# Multitasking Programming of OBDH Satellite Based On PC-104

Haryono
Indonesian National Institute of Aeronautics and Space (LAPAN)
Jl. Cagak Satelit KM.04 Rancabungur – Bogor 16310

*Abstract*— On Board Data Handling (OBDH) has functions to monitor, control, acquire, analyze, take a decision, and execute the command. OBDH should organize the task between sub system. OBDH like a heart which has a vital function. Because the function is seriously important therefore designing and implementing the OBDH should be carefully, in order to have a good reliability. Many OBDHs have been made to support the satellite mission using primitive programming. In handling the data from various input, OBDH should always be available to all sub systems, when the tasks are many, it is not easy to program using primitive programming. Sometimes the data become corrupt because the data which come to the OBDH is in the same time. Therefore it is required to have a way to handle the data safely and also easy in programming perspective.

In this research, OBDH is programmed using multi tasking programming perspective has been created. The Operating System (OS) has been implemented so that can run the tasks simultaneously. The OS is prepared by configuring the Linux Kernel for the specific processor, creating Root File System (RFS), installing the BusyBox. In order to do the above method, preparing the environment in our machine has been done, they are installing the Cross Tool Chain, U-Boot, GNU-Linux Kernel Source etc. After that, programming using c code with multitasking programming can be implemented.

By using above method, it is found that programming is easier and the corruption data because of reentrancy can be minimized.

## I. INTRODUCTION

National Institute of Aeronautics and Space (LAPAN) - Satellite Technology Center has main task to research and develop the satellite technology. One of the research is developing the On Board Data Handling (OBDH). OBDH which was developed for LAPAN satellites using primitive programming [1]. It is not easy to handle many input data when the data which have to be handled is more [2]. It is possible to be corrupted when the OBDH need to handle the data from many sub systems which come in the same time.

In the future LAPAN will create small satellite [3], many tasks will need to be handled by the OBDH. Therefore in order to support those requirement, new programming development need to be implemented. Primitive programming cannot handle the task simultaneously, unlike primitive programming, Real Time **Operating System** (RTOS) can handle multi tasks in parallel manner [4]. Primitive programming works based on interruption, when interruption is occurred in the same time, it is possible the data become corrupt or cut. In primitive programming, to minimize the data corruption we need to handle the task as fast as possible, if some tasks require a lot of computation, it is not easy to achieve that. In term of programming perspective it is not easy to program many tasks because the code is executed in sequent manner.

To handle the above problem, multi tasking programming has been implemented. By using multi tasking programming it is easy to program each task accordingly. In this research has main objective to develop the OBDH in the form of multi tasking programming. So that concept operation and mission can be fulfill by the OBDH easily.

## II. METHODS

### II.1. Methodology

Figure 1 is the methodology to achieve the aim of this research. Below is the step of the research activities which have been done:

1. Requirement Analysis: All the requirement of the system in the phase will be defined.
2. System Design: Define the hardware and software specification, including the specification of the hardware which has space qualification and programming method for OBDH.
3. Implementation: Preparing the environment and programming the OBDH as the requirement.
4. Integration and Testing: Integrate the system and create unit test each task, and the qualification test.
5. Analysis: Analysis the result and compare with other research that have been done.

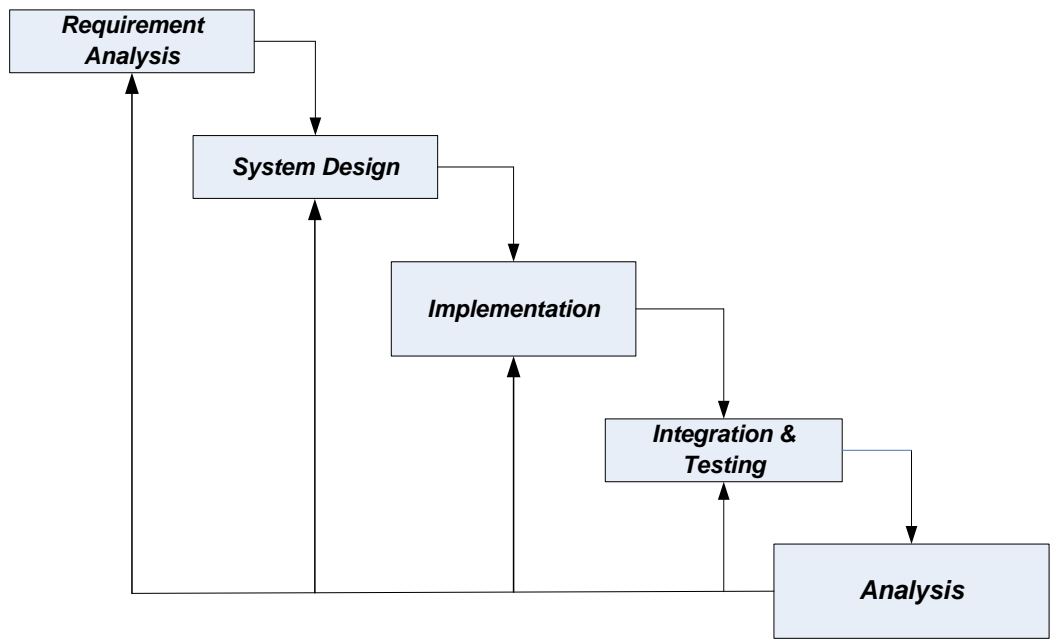

Figure 1. Methodology

### II.2. Requirement Analysis

In this phase will be described the satellite requirement to the OBDH. In the starting point, the OBDH will handle the Wheel Drive Electronics (WDE), Star Sensor (STS) and other sub systems. WDE will be 4 items, each should be handled simultaneously and STS will be 1 item. Wheel Drive Electronics (WDE) has TTL protocol and STS is RS422 protocol. The OBDH should also save the STS data to the memory safely. All tasks should be run in multitasking perspective.

### II.3. System Design Of On Board Data Handling (OBDH)

**Hardware Design**: in order to make fast in the development, **PC-104** has been selected, because of its known heritage design in space application [5] [6] [7] [8]. **PC-104** board which is selected should be space qualified.

Main board of the OBDH is **MPL-MIP405-3**, in this board our **operating system** is run and our programmed is implemented.

**Software Design:** the software which is designed using **Operating System** (OS), by using OS it easy to implement the multi tasking programming. Below are the advantages when using multi tasking programming:

1. Each task can run simultaneously [9].
2. Interruption will not disturb the task which is run.
3. From programming perspective is more easy to classify the task. In the line code execution is not done in sequential manner but parallel.

### II.4. Proposed Research

To make a better explanation of Proposed research, a schematic diagram is shown in the figure 2. The research focus is to implement the multitasking programming and to prove that multitasking programming has many advantages over the primitive programming.

In order to support the OS to be working, board selection was done. Many **Operating system** that is available in the market, from free until paid OS, this work part need to be considered according to the resources and the budget. After OS has been chosen, developing the OS and implementing to the board. Then the application need to be programming by using **C Programming**. Analysis and Result is important part, to see the result and to conclude about the finding whether successfully or not.

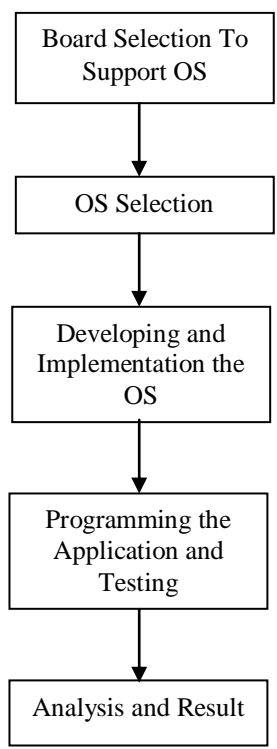

Figure 2 Schematic diagram of proposed research.

### III. IMPLEMENTATION

In order to implement the objective, two kind environment need to be prepared. Those two kinds are described below:

**1. Development Host Environment:**
a. Linux **Operating system** (Fedora)
b. Eclipse Development Tool
c. TFTP, SCP and SSH
d. Installation of Cross tool chain, U-Boot, Busybox, Linux **Kernel** modules, Creating and populating RFS, install SCP, GDB Server, OSCI driver etc.
e. Create the image disk for the target device.
f. The hardest part to provide the environment above is to prepare at the fourth step, it is required to prepare/install all the requirements before we prepare for the fourth step.

**2. Target Environment:**
MPL MIP405 is used to run the image which has been created from the host environment. It has 400 MHz processor and 8 MB flash memory [10].

### III.1. Programming Environment

In order to program the target, it is needed to prepare the environment, starting in creating the executable file until making the debugging. Executable file is built from **gcc** compiler. Linux **Kernel** is used as the **operating system** for the target device; hence the executable file should able to be run in Linux **operating system** environment.

MPL MIP405 is an IBM computer, Power PC as the core of the chip. In creating the executable, compiling the code should complain with the Power PC 405 (**PPC405**) otherwise the executable file will not run in the target device. In order to compile the code to executable file for PPC405, Installation of Cross tool chain in development tool is necessary. In this research Fedora Linux is selected to host the compiler for the PPC405. After preparing the compiler, the next step is to prepare the debugging environment. There are many ways to do this debugging, in term of communication and software that is used. Communication which is used when debugging is done in the Development Host computer most uses TCP/IP, Serial Port can be done but relatively slow. If the target device is complete device and has a good spec computer, developing in the target device directly is advice.

But when the target device is very minimum device then creating **Development Host** is needed. In this scenario second method is chosen. Figure 3. is a scenario to develop the target device.

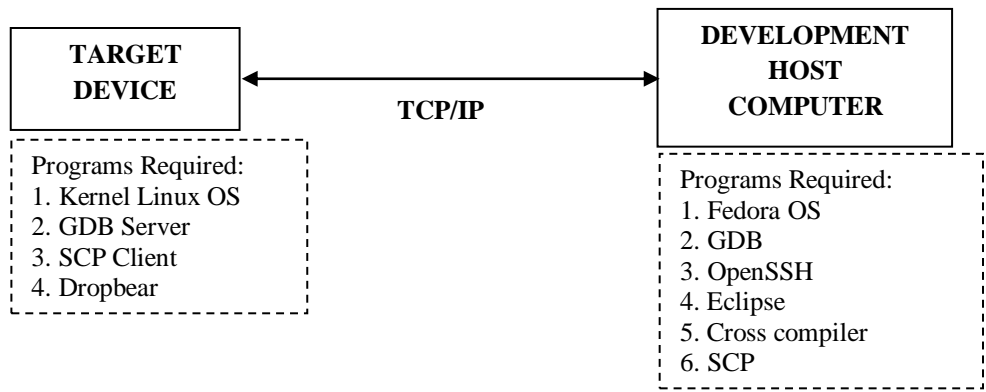

Figure 3. Scenario to develop the target device

Code is written in the Development Host Computer (DHC) via Eclipse IDE. Eclipse IDE with the Cross Compiler will work together to create the executable file. After executable file is created, it is needed to transfer via TCP/IP trough SCP. SCP is having task to transfer the file manually from DHC to target device. After the file is transferred to the target device, it is ready to run the file via GDB Server. In order to enable the debugging process, run the file via GDB server is required. In the eclipse IDE configuration to enable the debugging process was done. The parameter configuration is about the TCP/IP of the target device where the GDB Server is take place and location file of the executable in the target device. Before that installation of GDB server and Dropbear in target device are required. GDB server will execute the executable in target device, whereas the Dropbear will handle the communication with the DHC. Debugging communication is using SSH. Because the target device is limited resource to replace the OpenSSH, Dropbear is selected. Dropbear is simple and compact to support communication via SSH, versatile for the embedded device, small SSH server and client [11]. After those step have been prepared, debugging of the device over eclipse IDE via SSH can be done.

Note: When target device is not installed the OpenSSH, it is not possible to transfer the file directly via Eclipse. There is configuration in eclipse to disable the automatic transfer. Disable the automatic transfer is required, otherwise error will be show up and break up the process.

**III.2. Preparing the Kernel Linux OS**

Many **operating systems** can run in the MPL405 from free OS to License OS. The license OS like QNX or WindRiver is relatively expansive. It costs around $ 134000. In compensation of cost, it has a lot of resource to program the device, easy to prepare the debugging environment and have fast boot up. Because in our institution has limitation budget free OS is chosen. We don't use any distribution of Linux, but creating the image OS via **kernel** is done. The advantage using **kernel** Linux directly can minimized the image file as minimum as possible according to the need of the requirement and has ability to configure the OS as needed.

The step to create the Image OS for the target device is below:

[1] Download the zip file of the **kernel**
[2] Install the Cross Compiler for the PPC405
[3] Configure the **Kernel**
[4] Install the **Kernel**
[5] Install Busy box
[6] Create the Root File System (RFS)
[7] Extract the **Kernel** and Busy box to RFS
[8] Convert the RFS + **Kernel** to the image file system
[9] Burn the image file system to the target device

The image file system is binary file that can be copied to the non volatile memory in MPL405 via u-boot. U-Boot is open source, it has task to boot up the MPL405 device. After the device is boot up, the image file can be copied to the device via Serial port or LAN.

**III.3. Starting the Development Environment**
**1st Step, Preparing On the Target:**

After OS image has been created and copied to the target flash, the target is ready to be run. *bootm* is the boot command to run the OS image that was copied. The prompt will show up, need to login as **root** with appropriate password. Serial Port and Local Area Network should be connected to the development PC.

**2nd Step, Preparing On the Development PC:**

In normal PC there is monitor it can be essay to command the Linux Box and see the result, but for MPL405 there is no VGA card to show the console window to the screen. So that serial communication to communicate between the target and the development PC is used. Hyperterminal and appropriate setting has been applied. Serial communication setting is 9600 8 N 1. Figure 4 is about the command to the target in order to start the environment development. Before to start the *dropbear*, *dropbear* has been installed to the OS. *dropbear* task is like SSH, the main task of *dropbear* is copying the executable program from development PC to the target machine. *modprobe mpl_osci.ko* is command to start up the driver for the serial communication board (extended **PC-104** Board). It has 8 ports each board, can be operated as RS232 or RS 422 or TTL. The next command is about the setting for the port to be RS 422/RS232/TTL.

```
mkdir /etc/dropbear
cd /etc/dropbear
dropbearkey -t dss -f dropbear_dss_host_key
dropbearkey -t rsa -f dropbear_rsa_host_key
dropbear -s -g

modprobe mpl_osci.ko
/mnt/hda1/set_osci_hryn /dev/ttyOS0 -r 4 -f 1
/mnt/hda1/set_osci_hryn /dev/ttyOS1 -r 4 -f 1
/mnt/hda1/set_osci_hryn /dev/ttyOS2 -r 4 -f 1
/mnt/hda1/set_osci_hryn /dev/ttyOS4 -r 4 -f 1
/mnt/hda1/set_osci_hryn /dev/ttyOS6 -r 4 -f 1
```

Figure 4. Command to start the development environment in the target

**3th Step:** In this step will be discussed about the programming and debugging of the target using Eclipse. Below is the step:

**- Start Fedora 21:** Many Distros have been tried to create the development PC, but end up with some failing in the installation. When come to Fedora 21 development PC can be prepared smoothly.

**- Run Eclipse:** By using eclipse debugging is more easy and free software. Eclipse acts as integrated development environment. **C programming** is selected to program the target.

**- Preparing the Executable File in the target:** To get executable file, C code that has been write up can be compiled. The compiler that has been used is *Crosstool NG*. *Crosstool NG* at here acts to compile the code to the MPL405, it is using PPC 405 processor. *Crosstool NG* has been installed successfully in Fedora 21 for PPC 405 processor.

By pressing the Ctrl + B get the executable file for the target. Executable file is sent via *Dropbear* using command: **scp exc1 root@192.168.137.2:/home**, means copy the **exc1** file to the target device with IP Address 192.168.137.2 in **home** directory. Need to make sure the **ssh** communication between target and development PC is established.

## IV. RESULTS AND ANALYSIS

**IV.1. Testing the software**

In this topic will discuss about the testing of the OBDH based on multitasking. After everything is done in implementation the next test is to verify whether the OS and the software that is burn in to the **PC-104** is working properly.

The test is done by two categories:
1. To know whether OS work and all tasks can run multitasking.
2. To know each task is working properly as assigned.

Category one, to know whether OS work and all tasks can run multitasking, in figure 5 is shown to describe that OS have run smoothly. It is shown that Linux OS is run and gave us a prompt and ready to be commanded. Each task is executed, one it is executed will run and ready to receive the data input from Receiver port (Rx).

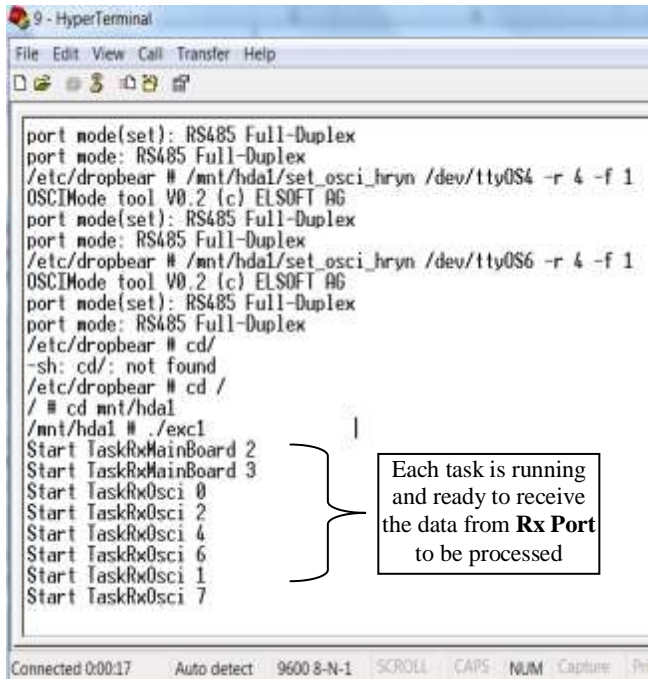

Figure 5. Linux OS has run and executed each task

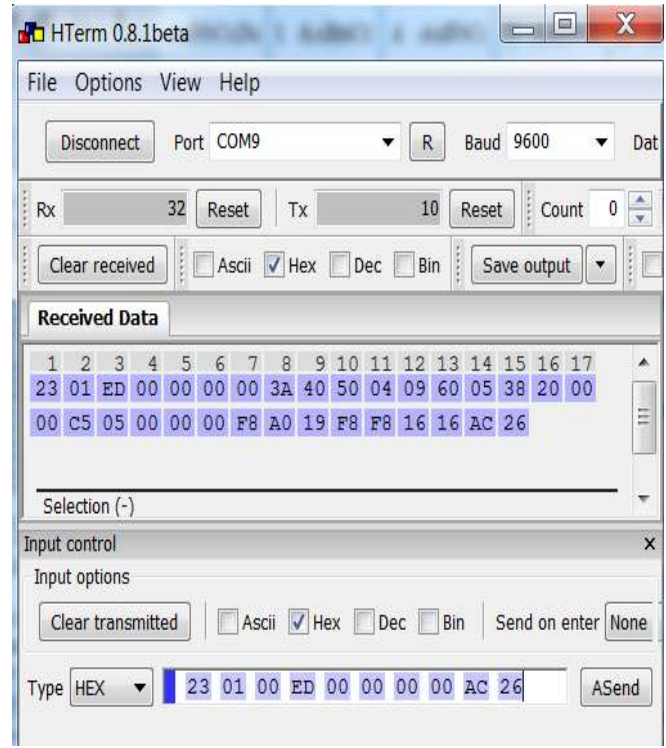

Figure 6. Command to OBDH to get System Telemetry from WDE

Category two, to know each task is working properly as assigned, in figure 6 showed that the software has been tested by commanding the OS to get the Wheel Drive Electronic (WDE) data. The OBDH reply back and give the WDE data correctly, the data is showed in Receive Data column with 32 byte data. The testing to the sub system has been done and applied to all sub systems as shown in Table 1 and the result was getting the correctly data. All communication to sub system is done in multitasking perspective.

| Device ID | Name |
|---|---|
| 1 | WDE 1 |
| 2 | WDE 2 |
| 3 | WDE 3 |
| 4 | Start Sensor (STS) 1 |
| 5 | Start Sensor (STS) 2 |
| 6 | BATREY (Simulation) |
| 7 | GPS (Simulation) |

Table 1. Sub systems that have been connected and tested to communicate with the OBDH

## IV.2. Multi Tasking Programming perspective

In this topic will discuss about differences programming perspective between primitive programming and multi tasking programming. In Figure 7 (primitive programming) and Figure 8 (multi tasking programming) are a **C programming** that have been implemented in the microcontroller and microprocessor.

```
main ()
{
        TaskRxMainBoard2();
        TaskRxMainBoard3();
        TaskRxOsci0();
        TaskRxOsci2();
        TaskRxOsci4();
}
```

Figure 7. Primitive programming

```
void* TaskRxMainBoard2(void *arg)
{
        // Task code
}
void* TaskRxMainBoard3(void *arg)
{
        // Task code
}
void* TaskRxOsci0(void *arg)
{
        // Task code
}
void* TaskRxOsci2(void *arg)
{
        // Task code
}
void* TaskRxOsci4(void *arg)
{
        // Task code
}
```

Figure 8. Multi tasking programming

In Figure 7. the code is executed sequence, it is not easy to fulfill the requirement of the concept operation. It can be solved by the interruption technique but when requirement is more complex, sometime the task that is running can be interrupted by interruption immediately. When using multitasking in Figure 8, it can be seen that programming is more easy. Each of task run simultaneously, task can be sleep, suspended or run according to the need. Without any dependency to other tasks, when the task is completed in their process.

**IV.3. Non Reentrancy Function**

In primitive method, it is possible to get corrupt data because neglecting the non Reentrancy function. As figured in the figure 9, it showed that the data being corrupted because in the same time using same non Reentrancy function, the process is interrupted while sending the data. As shown in the figure 9, stream data of lowercase alphabet is cut by uppercase alphabet.

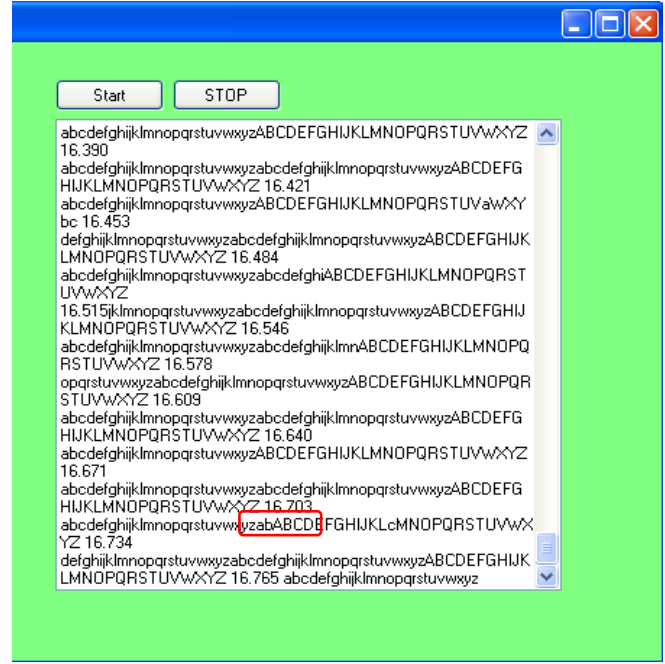

Figure 9. Primitive method faces the corrupt data [12].

**IV.4. Portable Operating System Interface (POSIX) standard**

Embedding Linux OS in the target device will have ability in using the standard POSIX coding. A lot of resources or libraries that can be used to create a code based on POSIX standard. Because of that the code that has been created can be more adaptable and easier to be created.

**IV. CONCLUSION AND FUTURE WORK**

The conclusion that can be obtained is by using Multi Tasking Programming perspective, it is found that programming is easier and the corruption data because of reentrancy can be minimized.

The next research will implement this PC-104 board with operating system for On Board Data Handling sub system of LAPAN Satellites. It will be further analyzed in to the real application of On Board Data Handling.

**ACKNOWLEDGMENT**

This project is supported by Satellite Centre - LAPAN, **Mr Abdul Karim** as Head of Bus Satellite LAPAN, **Mr Suhermanto** as Head of Satellite Center LAPAN and my colleges Mr Fauzi and Mr Taufik. We would like to acknowledge for their support in this project.